\documentclass[aps,twocolumn,prd,preprintnumbers, groupedaddress, nofootinbib, 10pt]{revtex4-1}
\pdfoutput=1
\usepackage[hidelinks]{hyperref}
\usepackage{amssymb}
\usepackage{amsfonts}
\usepackage{graphicx}
\usepackage{epstopdf}
\usepackage{dcolumn}
\usepackage{amsmath}
\usepackage{latexsym,bm}
\usepackage{amsthm}
\usepackage{slashed}
\usepackage{float}
\usepackage{color}
\usepackage{url}
\usepackage{longtable}
\usepackage{rotating}
\usepackage[normalem]{ulem}
\usepackage{faktor}
\usepackage{tikz-cd}
\usepackage{tensor}

\newcommand{\lb}{\left(}
\newcommand{\rb}{\right)}

\newcommand{\half}{\frac{1}{2}}

\newcommand{\Z}{{\mathbb Z}}

\newcommand{\cK}{\mathcal{K}}
\newcommand{\cL}{\mathcal{L}}
\newcommand{\cM}{\mathcal{M}}
\newcommand{\cN}{\mathcal{N}}

\newcommand{\cR}{\mathcal{R}}

\renewcommand{\d}{\text{d}}

\newcommand{\be}{\begin{equation}}
\newcommand{\ee}{\end{equation}}
\newcommand{\ba}{\begin{aligned}}
\newcommand{\ea}{\end{aligned}}

\newcommand{\bea}{\begin{eqnarray}}
\newcommand{\eea}{\end{eqnarray}}

\def\half{{\frac{1}{2}}}

\def\unit{{1\kern-.65ex {\rm l}}}
\def\1{{1\kern-.65ex {\rm l}}}

\usepackage{verbatim}
\usepackage{tikz}
\usetikzlibrary{arrows,snakes,shapes.arrows,decorations.markings}
     \tikzset{>=triangle 90}
     \tikzstyle{gr}=[draw,circle,green!50!black,fill=green!50!black,scale=.6]
     \tikzstyle{Bl}=[draw,circle,blue,scale=.7]
     \tikzstyle{R}=[draw,circle,fill=red,scale=.7]
     \tikzstyle{bl}=[draw,circle,fill=black,scale=.2]
     \tikzstyle{bbc}=[draw,circle,fill=black,scale=.75]
     \tikzstyle{bbcs}=[draw,circle,fill=black,scale=.5]
     \tikzstyle{rc}=[circle,fill=red,scale=.6]
     \tikzstyle{wc}=[draw,circle,scale=.75]
\usepackage{array} 
\usepackage{multirow} 
\usepackage{colortbl} 
\usepackage{stfloats}  
\usepackage{cellspace}
\usepackage{xcolor}   

\newcommand{\xdasharrow}[2][->]{
\tikz[baseline=-\the\dimexpr\fontdimen22\textfont2\relax]{
\node[anchor=south,font=\scriptsize, inner ysep=1.5pt,outer xsep=2.2pt](x){#2};
\draw[shorten <=3.4pt,shorten >=3.4pt,dashed,#1](x.south west)--(x.south east);
}
}

\def\hat{\widehat}

\def\^{\wedge}

\def\Z{\mathbb{Z}}

\def\cK{{\mathcal K}}
\def\cM{{\mathcal M}}
\def\cN{{\mathcal N}}

\def\cR{{\mathcal R}}

\newcount\hour \newcount\minute
\hour=\time \divide \hour by 60
\minute=\time
\def\now{%
\ifnum \hour<13
  \ifnum \hour=0 \advance \hour by 12 \number\hour:\else \number\hour:\fi%
     \ifnum \minute<10 0\fi%
     \number\minute%
\ A.M.%
\else \advance \hour by -12 \number\hour:%
  \ifnum \minute<10 0\fi%
  \number\minute%
  \ P.M.%
\fi%
}

\usepackage{tikz}
\usetikzlibrary{arrows}
\usetikzlibrary{shapes.geometric,calc,arrows, positioning,shapes.misc,decorations.markings}
\tikzset{
  big arrow/.style={
    decoration={markings,mark=at position 1 with {\arrow[scale=2,#1]{>}}},
    postaction={decorate},
    shorten >=0.4pt},
  big arrow/.default=black}
\usetikzlibrary{external}
\usetikzlibrary{positioning}
\usetikzlibrary{calc}
\tikzset{gauge-node/.style={shape=circle, draw, minimum width=.6cm}}

\pgfdeclarelayer{edgelayer}
\pgfdeclarelayer{nodelayer}
\pgfsetlayers{edgelayer,nodelayer,main} 
\tikzstyle{none}=[inner sep=0pt] 

\tikzstyle{NodeCross}=[draw, shape=circle, cross out, inner sep=0pt, minimum size=6pt,line width=0.25mm]
\tikzstyle{Circle}=[draw, shape=circle, black, inner sep=0pt, minimum size=6pt]
\tikzstyle{rtriangle}=[fill=black, regular polygon, regular polygon sides=3, rotate=90, inner sep=0pt, minimum size=8pt]
\tikzstyle{ltriangle}=[fill=black, regular polygon, regular polygon sides=3, rotate=270, inner sep=0pt, minimum size=8pt]
\tikzstyle{rtriangleblue}=[fill={rgb,255: red,17; green,160; blue,255}, regular polygon, regular polygon sides=3, rotate=90, inner sep=0pt, minimum size=8pt]
\tikzstyle{ltriangleblue}=[fill={rgb,255: red,17; green,160; blue,255}, regular polygon, regular polygon sides=3, rotate=270, inner sep=0pt, minimum size=8pt]
\tikzstyle{rtrianglegreen}=[fill={rgb,255: red,69; green,255; blue,28}, regular polygon, regular polygon sides=3, rotate=90, inner sep=0pt, minimum size=8pt]
\tikzstyle{ltrianglegreen}=[fill={rgb,255: red,69; green,255; blue,28}, regular polygon, regular polygon sides=3, rotate=270, inner sep=0pt, minimum size=8pt]
\tikzstyle{Uprtriangle}=[fill=black, regular polygon, regular polygon sides=3, rotate=0, inner sep=0pt, minimum size=8pt]
\tikzstyle{Downltriangle}=[fill=black, regular polygon, regular polygon sides=3, rotate=180, inner sep=0pt, minimum size=8pt]
\tikzstyle{rtriangleAmber}=[fill={rgb,255: red, 191; green, 144; blue, 63}, regular polygon, regular polygon sides=3, rotate=90, inner sep=0pt, minimum size=8pt]
\tikzstyle{UprtriangleViolett}=[fill={rgb,255: red,255; green,0; blue,0}, regular polygon, regular polygon sides=3, rotate=0, inner sep=0pt, minimum size=8pt]

\tikzstyle{Downltriangle}=[fill=black, regular polygon, regular polygon sides=3, rotate=180, inner sep=0pt, minimum size=8pt]
\tikzstyle{UpRighttriangle}=[fill=black, regular polygon, regular polygon sides=3, rotate=45, inner sep=0pt, minimum size=8pt]
\tikzstyle{UpLefttriangle}=[fill=black, regular polygon, regular polygon sides=3, rotate=315, inner sep=0pt, minimum size=8pt]
\tikzstyle{DownRighttriangle}=[fill=black, regular polygon, regular polygon sides=3, rotate=135, inner sep=0pt, minimum size=8pt]
\tikzstyle{DownLighttriangle}=[fill=black, regular polygon, regular polygon sides=3, rotate=225, inner sep=0pt, minimum size=8pt]

\tikzstyle{Star}=[draw, shape=star, fill=black, star points=8, inner sep=0pt, minimum size=8pt]

\tikzstyle{DashedLine}=[-, densely dashed, line width=0.25mm]
\tikzstyle{DashedLineBrown}=[-, densely dashed, line width=0.25mm, draw={rgb,255: red,155; green,103; blue,51}]
\tikzstyle{DashedLineFall}=[-, densely dashed, line width=0.25mm, draw={rgb,255: red,195; green,0; blue,0}]
\tikzstyle{DashedLineViolett}=[-, densely dashed, line width=0.25mm, draw={rgb,255: red,139; green,41; blue,148}]
\tikzstyle{DottedLine}=[-, dotted, line width=0.25mm]
\tikzstyle{BlueLine}=[-, fill=none, draw={rgb,255: red,17; green,160; blue,255}, line width=0.25mm]
\tikzstyle{GreenLine}=[-, fill=none, draw={rgb,255: red,69; green,255; blue,28}, line width=0.25mm]
\tikzstyle{RedLine}=[-, draw={rgb,255: red,191; green,0; blue,0}, fill=none, line width=0.25mm]
\tikzstyle{DashedLineRed}=[-, densely dashed, fill=none, draw={rgb,255: red,191; green,0; blue,0}, line width=0.25mm]
\tikzstyle{ThickLine}=[-, line width=0.25mm]
\tikzstyle{ViolettLine}=[-, draw={rgb,255: red,132; green,60; blue,191}, fill=none, line width=0.25mm]
\tikzstyle{ViolettDashedLine}=[-, densely dashed, draw={rgb,255: red,132; green,60; blue,191}, fill=none, line width=0.25mm]
\tikzstyle{AmberLine}=[-, draw={rgb,255: red,191; green,144; blue,63}, fill=none, line width=0.25mm]
\tikzstyle{DashedRedThick}=[-, densely dashed, fill=none, draw={rgb,255: red,191; green,0; blue,0}, line width=0.40mm]
\tikzstyle{DashedBlueThick}=[-, densely dashed, fill=none, black, line width=0.40mm]

\makeatother

\begin{document}

\title{Holography, 1-Form Symmetries, and Confinement}

\author{Fabio Apruzzi}
\author{Marieke van Beest}
\author{Dewi S.W.~Gould}
\author{Sakura Sch\"afer-Nameki}

\affiliation{Mathematical Institute, University
of Oxford, Woodstock Road, Oxford, OX2 6GG, United Kingdom}


\begin{abstract}
\noindent 
We study confinement in 4d $\mathcal{N}=1$ $SU(N)$ Super-Yang Mills (SYM) from a holographic point of view, focusing on the 1-form symmetry and its relation to chiral symmetry breaking. In the 5d supergravity dual, obtained by truncation of the Klebanov-Strassler solution, we identify the topological couplings that determine the 1-form symmetry and its 't Hooft anomalies. One such coupling is a mixed 0-form/1-form symmetry anomaly closely related to chiral symmetry breaking in gapped confining vacua. From the dual gravity description we also identify the infra-red (IR) 4d topological field theory (TQFT), which realises chiral symmetry breaking and matches the mixed anomaly. Finally, complementing this, we derive the chiral and mixed anomalies from  the Little String Theory realization of pure SYM. 

\end{abstract}



\maketitle


\noindent
\textbf{Introduction.}
Global symmetries and their 't Hooft anomalies can highly constrain the dynamics of gauge theories. 
A prime example is the role of the 1-form symmetry in  confinement of $\mathcal{N}=1$ $SU(N)$ super Yang-Mills (SYM) or adjoint QCD theories. In this case the 1-form symmetry $\Gamma^{(1)}=\mathbb{Z}_N$ and corresponds to the center of the gauge group, which acts on line operators \cite{Aharony:2013hda, Gaiotto:2014kfa} and provides a diagnostic of confinement. The order parameter for this symmetry is the vacuum expectation value (vev) of the Wilson line in the fundamental representation, which obeys  area law in a confining vacuum. This implies that an infinitely extended Wilson line has vanishing vev, therefore preserving the 1-form symmetry. 
In addition, $\mathcal{N}=1$ $SU(N)$ SYM also has a 0-form R-symmetry $U(1)_R^{(0)}$. 
The  Adler-Bell-Jackiw (ABJ) or chiral anomaly breaks $U(1)_R^{(0)}$ to  $\Gamma^{(0)}= \mathbb{Z}_{2N}$, which by chiral symmetry breaking \cite{Witten:1982df} ($\chi_{\text{SB}}$) further breaks to $\mathbb{Z}_2$ in the confining phase
\be\label{BreakingStuff}
U(1)\quad  \stackrel{\text{ABJ}}{\longrightarrow} \quad \mathbb{Z}_{2N}\quad  \stackrel{\chi_\text{SB}}{\longrightarrow} \quad \mathbb{Z}_2 \,.
\ee
There is a 0-/1-form symmetry mixed 't Hooft anomaly
\be\label{COanomaly}
\mathcal{A}[b_2, A] = 2\pi \, N^2 \int_{X_5} A \, b_2 \, b_2 \,,
\ee
where $b_2$ is the background for $\mathbb{Z}_{N}^{(1)}$ and $A$ for $\Gamma^{(0)}$, which satisfy 
$\oint b_2 \in \frac{\mathbb{Z}}{N}$ and $\oint A \in \frac{\mathbb{Z}}{2N}$. This anomaly constrains the IR strongly coupled physics   \cite{Gaiotto:2014kfa, Cordova:2018acb, Cordova:2019bsd}. 
In a confining vacuum the 1-form symmetry is unbroken, and the 0-form background has to satisfy $\oint  A \in \frac{\mathbb{Z}}{2}$ and $\Gamma^{(0)}$ is broken to 
$\Gamma^{(0)}=\mathbb{Z}_2$. This breaking indicates $N$ distinct confining vacua, modelled by a gapped TQFT.

The goal of this paper is to take a holographic perspective, from which we derive the 1-form symmetry and the mixed anomaly, as well as the TQFT describing the IR confining vacua. 
Higher-form symmetries in the AdS/CFT correspondence were discussed in \cite{Witten:1998wy, Aharony:1998qu, Gross:1998gk, Hofman:2017vwr, Bergman:2020ifi}. Our focus here is on holography in a non-conformal setting, where the dual gauge theory is conjectured to be a confining theory related to $\mathcal{N}=1$ $SU(N)$ SYM \cite{Seiberg:1994rs,Seiberg:1994aj,Klemm:1994qs,Argyres:1994xh}. Concretely, we develop the methods to determine the 1-form symmetry in the Klebanov-Strassler (KS) \cite{Klebanov:1999rd,Klebanov:2000hb, Klebanov:2000nc, Herzog:2001xk, Klebanov:2002gr, Strassler:2005qs} solution. The central tool for our analysis is the consistent truncation of supergravity to 5d \cite{Cassani:2010na, Bena:2010pr, Schon:2006kz}. From this we determine a St\"uckelberg coupling for the R-symmetry which breaks it to a discrete subgroup as predicted by the ABJ anomaly in field theory, as well as 5d topological couplings from which we identify the 1-form symmetry and anomalies that will be central to $\chi_{\text{SB}}$. Finally we show how the 5d supergravity reduction contains as boundary counterterms the 4d TQFT describing the IR vacua of $\mathcal{N}=1$ $SU(N)$ SYM.
The approach proposed in this paper has a vast number of generalizations, to holographic setups for confining theories, but also to geometric engineering constructions of confining theories e.g. \cite{Atiyah:2000zz}. It provides an exciting opportunity to revisit these setups, and sharpen the predictions, by taking the perspective based on higher-form symmetries.

\vspace{0.1cm}
\noindent
\textbf{Holographic Dual to Confinement.}
One of the most successful holographic realizations of $\mathcal{N} =1$ $SU(N)$ SYM theory is the KS-solution \cite{Klebanov:2000hb}. This construction is realised in 10d IIB  supergravity, and it consists of two main ingredients: (1) $N$ D3-branes probing the conifold $C(T^{1,1})$, which is a conical Calabi-Yau with 5d link $T^{1,1}$, that is topologically $S^2 \times S^3$. The near-horizon of this brane system is AdS$_5 \times T^{1,1}$ with 5-form flux $\int_{T^{1,1}} F_5 =N$.
(2) $M$ D5-branes wrapping the $S^2 \subset T^{1,1}$. The D5s backreact on the external geometry, modifying the AdS$_5$ metric.  The solution at large radial distances,  the KS-solution, is
\be\ba \label{eq:conmet}
\d s^2_{10} &=\d s^2_{\cM_5} + \cR^2(r)\d s^2_{T^{1,1}}\,, \quad \cR(r) \sim \text{ln}\left({r \over r_s}\right)^{1/4}\,,
\ea\ee
where $\d s^2_{\cM_5}=\frac{r^2  \d \vec{x}^2}{\cR^2(r)}+{\cR^2(r) \d r^2  \over r^2} $, and  $r_s =r_0 e^{ -\frac{2 \pi N}{3 g_s M^2} -\frac{1}{4}}$.
At large $r$, the quantization of fluxes is 
\be\ba
\int_{S^{3}} F_{3}  &= M\,, \quad  
 \int_{S^{2}} B_{2}  = \cL(r) \,, \\
\int_{T^{1,1}}F_{5} &=\cK(r) = N+M \cL\,, \quad 
\mathcal{L}= \frac{3g_{s}M}{2\pi} \text{ln}(r/r_{0})\,,
\ea\ee
where $r_0$ is the UV scale, and we refer to this as the UV KS-solution, valid for $r$ sufficiently large and $g_s\cK(r) \gg 1$.
Note that $F_5$ is no longer quantised: its integral over the internal space acquires a radial dependence. 
The solution has a naked singularity at $\cR(r_s)=0$, and in particular we can consider $r_s \rightarrow 0$, when $\frac{N}{M^2} \gg 1$. 
Due to the naked singularity at small radial distances $r \rightarrow r_s$, higher curvature corrections become relevant, and \eqref{eq:conmet} is no longer valid. There is a smooth solution describing this regime, and it requires the full warped, deformed conifold solution \cite{Klebanov:2000hb}, see appendix \ref{app:FullSol}.

The dual field theory description is given by $SU(N+M) \times SU(N)$ gauge theory and bifundamental matter in $(\textbf{N}+{\bf M}, \overline{\textbf{N}}) \oplus (\overline{\textbf{N}+{\bf M}}, \textbf{N})$, where a combination of the two gauge couplings has flown to strong coupling regimes. In particular, this theory is not conformal and the gauge couplings of the two factors run in opposite directions. E.g. when $SU(N+M)$ with $N_F= 2N$ becomes strongly coupled, we apply Seiberg duality \cite{Seiberg:1994pq}, resulting in $SU(N-M)$ with $N_F= 2N$. This process perpetuates with the new gauge couplings flowing in opposite directions, giving rise to a `duality cascade'. For $N=\kappa M$, $\kappa\in \mathbb{N}$, the endpoint is $\mathcal{N} =1$ $SU(M)$ SYM at strong coupling. 

The RG-flow of the gauge theory cascade is mirrored explicitly in the dual gravity background. Moving from large $r$ to $r \to re^{-\frac{2\pi}{3g_s M}}$, 
 $\int_{S^2}B_2$ and $\int_{T^{1,1}} F_5$ change by $\cL(r) \to \cL (r)-1, \; \cK(r) \to \cK(r)-M$. 
At the special slices with fixed $r=r_k= r_0 e^{-\frac{2\pi k}{3g_s M}}$,
where  $\cL, \cK$ are integer, the supergravity background is dual to the  $SU(N-(k-1)M) \times SU(N-kM)$ gauge theory in the baryonic branch \cite{Dymarsky:2005xt}. {Alternatively we can work in terms of Page charges defined in \cite{Benini:2007gx}, where $\hat{F}_5= F_5 - B_2 F_3$ is always integrally quantized  $\int_{T^{1,1}} \hat{F}_5 = N- kM$, due to large gauge transformations of $\int_{S^2} B_2 \rightarrow \mathcal{L}(r) + k$.} For $N= \kappa M$ the endpoint is reached at a value $r_{\kappa}$ where there are only $M$ units of $F_3$-flux and no $F_5$-flux. In this regime, $r \sim r_{\kappa}$, the solution \eqref{eq:conmet} breaks down before reaching the $r \rightarrow r_s$ limit,  since $g_s\cK(r_{\kappa}) = 0$, and the metric in \eqref{eq:conmet} is not smooth. We therefore have the following hierarchy of scales $r_0 \gg r_k \gg r_{\kappa} > r_s$.
The smooth supergravity solution for $r\leq r_{\kappa}$ is instead provided by the warped deformed conifold \cite{Klebanov:2000hb}. We denote this by the IR KS-solution, which describes the IR regimes of $\mathcal{N} =1$ $SU(M)$ SYM. 

\vspace{0.1cm}
\noindent
\textbf{1-Form Symmetries from Supergravity.}
The global form of the gauge group, or put differently, the set of mutually local line operators, 
can be determined in holography by considering boundary conditions (b.c.) of Chern-Simons-like couplings \cite{Witten:1998wy, Aharony:1998qu, Gross:1998gk}. Put in a more modern language, the 2-form backgrounds for 1-form symmetries of the 
holographic field theory are determined by topological couplings in the bulk and specific b.c.s yield absolute theories (i.e. definite spectra of line operators). On the gauge theory side of the duality cascade, the $\mathbb{Z}_{N+M} \times \mathbb{Z}_{N}$ center symmetry of $SU(N+M)\times SU(N)$ is broken by the matter to 
\be\label{eq:1fs}
\Gamma^{(1)} = \mathbb{Z}_{\text{gcd}(N,N+M)} = \mathbb{Z}_{\text{gcd}(N, M)}\,.
\ee
This remains constant through each step of the cascade.

In order to derive the 1-form symmetry holographically we study fluctuations around the UV KS-solution, when $r$ is sufficiently large and $g_s\cK(r) \gg 1$, which describes each step of the cascade until $r\sim r_{\kappa}$. The latter corresponds to the end of the cascade and the smooth gravity dual is the warped deformed conifold, which we will investigate momentarily. 
We reduce IIB supergravity on $T^{1,1}$ and study the topological couplings of the 5d 2-form gauge fields on this background. The strings which couple to these fields induce line operators on the $r_k$ slices, which in turn furnish the 1-form symmetry of the boundary theory, also known as a `singleton theory' \cite{Belov:2004ht,Maldacena:2001ss}. 
In particular the 1-form symmetry is deduced from the bulk couplings by imposing a set of consistent b.c.s in which a subset of the 2-form gauge potentials are fixed in a subgroup of $U(1)$. We first consider fluctuations of the fluxes around the background on the cohomology of $T^{1,1}$, i.e. $F_q$ expanded along $\omega_p \in H^p(T^{1,1},\Z)$ as $F_q = \sum_p f_{d-p} \wedge \omega_p$. In this case the non-trivial cohomology elements are $\omega_2$ and $\omega_3$, i.e. the volume forms of $S^2$ and $S^3$ respectively. The fluctuations of the IIB fluxes around the KS-solution read
\be \label{eq:fluct}
\ba 
\delta H_3 &=db_2\,, \quad
 \delta F_3 = dc_2+ \omega_2 \wedge dc_0 \cr
\delta F_5 &= \omega_2 \wedge  f_3 +  \omega_3 \wedge \ast f_3\,,
\ea 
\ee 
where these expansions already satisfy the Bianchi identities for $H_3, F_3, F_5$ and self-duality $F_5=\ast F_5$  (see appendix \ref{app:eom}). Here, $c_0$ is a $2\pi$ periodic scalar, whereas $b_2, c_2$ couple to F1s and D1s, respectively. We use $f_3$, since the operators of the boundary 1-form symmetry are manifest in this frame. The  Bianchi identity for $F_5$ sets $f_3 - \cL dc_2 -b_2dc_0  =d a_2$, which introduces a 2-form gauge field sourced by D3s wrapping the $S^2 \subset T^{1,1}$.

In order to get the 5d bulk topological action describing the singleton theory, we implement the following strategy. We reduce the flux equations of motion of IIB supergravity on $T^{1,1}$ and construct a classical 5d action which realises these equations of motion, see appendix \ref{app:eom}.  We study this at $r=r_k+r'\,, \ r' \ll r_k$, which has a field theory dual.  For $k \ll \frac{N}{M}$ we are far away from the IR cut-off and in this regime the topological terms dominate. The effective $F_5$ charge is $\cK(r)= N-k M + \mathcal{O}\left(r'/r_k\right) $, so the leading contributions to the equations of motion are the topological couplings
\be 
\text{gcd}(N,M) d\mathcal{C}=0\,,  \qquad \text{gcd}(N,M) db_2 =0\,,
\ee 
where $\mathcal{C} = q_1c_2 - q_2a_2$, with  $\text{gcd}(N,M)q_1=N$, $\text{gcd}(N,M)q_2=M$, and whereby we decoupled the center of mass $U(1)^{(1)}$, corresponding to the 1-form symmetry of the collective motion of the D3s.
The couplings are embedded into the consistent truncation of \cite{Cassani:2010na}. One can compare by varying their topological action, changing duality frame and restricting to the relevant fields. 
We  find that the following topological term in the 5d supergravity reduction on the UV KS-solution at $r=r_k \gg r_{\kappa}$ dominates over higher derivative couplings, 
\be \label{eq:truetop}
S_{\text{5d}} \supset 2 \pi  \; \text{gcd}(N,M)\int_{\cM_5} b_2 \wedge d \mathcal{C}\, .
\ee

1-form symmetries are generated by \textit{topological} surface operators \cite{Gaiotto:2014kfa}, which here are 
$U_{b}(M_2) =e^{2 \pi i\oint_{M_2} b_2}$ and $U_{c}(M_2) =e^{2 \pi i\oint_{M_2} \mathcal{C}}$,
where $M_2$ are closed surfaces, $\partial M_2 = \emptyset$. Generically, due to non-commutativity of fluxes, these do not commute \cite{Witten:1998wy}
\be
U_b (M_2) U_c (N_2) =  U_c (N_2) U_b (M_2) e^{ 2 \pi i  L( M_2, N_2 )\over N}\, ,
\ee
where $L$ is the linking of the surfaces.  These \textit{charge} operators generate a 1-form symmetry, which acts on \textit{charged} line operators in the 4d field theory. We find these charged line operators by considering operators of the form $U_{b} (\Sigma)$ with $\partial\Sigma \subset \cM_5|_{r_k}$. Similarly, the line operators $U_b(\Sigma)$ and $U_c(\Sigma)$ are not mutually local due to their linking. At each $r_k$ slice, a maximal set of mutually local line operators corresponds to b.c.s of $b_2$ and $\mathcal C$. 

A possible choice of b.c. for \eqref{eq:truetop} is 
$b_2$  Dirichlet and $\mathcal{C}$ Neumann. 
Since $\mathcal{C}$ is free to vary at the boundary, $U_c$ will correspond to the topological charge operator for the 1-form symmetry. By varying the topological action we find a condition
$
\text{gcd}(N,M) b_2 \wedge \delta \mathcal{C} |_{r_k}= 0\,,
$
which forces $b_2$ to take fixed values at the boundary. This implies that we can define a flat connection $b_1$ in 4d taking values in $\mathbb{Z}_{\text{gcd}(N,M)}$, i.e. $\text{gcd}(N,M) b_2= db_1=0$ at the slice $r=r_k$.  Therefore, $U_b$ restricted to $\partial \Sigma \subset  \cM_5|_{r_k}$ corresponds to the charged line operators of the field theory. As is well known, the fundamental strings, carrying world-volume $b_2$, ending on the boundary indeed give rise to Wilson lines in the 4d theory. They generate the 1-form symmetry $\Gamma^{(1)}$ of \eqref{eq:1fs}.
The screening can equally be seen by considering the analog of the `baryon vertex' \cite{Witten:1998xy} in this setup: 
integrating the Bianchi identities for D5s on $T^{1,1}$ and D3s on $S^3$ yield
\be
\ba
\int_{T^{1,1}} dF_7 =\int_{T^{1,1}} H_3 \wedge F_5 =&\  (N -kM)H_3
 \cr
\int_{S^3} dF_5 =\int_{S^3} H_3 \wedge F_3  = &\ M H_3 \,.
\ea
\ee
Thus D5s on $T^{1,1}$ provide the `baryon vertex' that screens $N-kM$ F1s, while D3s on $S^3$  screen $M$ F1s. In particular, $\text{gcd} (N,M)$ F1s is the minimal configuration of strings that  are screened. 
Alternative b.c.s can be studied \cite{otherbc}, the IR will fix the one above.

From here onwards, we consider $N=\kappa M$, which allows us to connect to confinement. In this case 
the deformed conifold IR solution holographically describes the bottom of the cascade for the confining phase of SYM with $SU(M)$ simply-connected gauge group. We find $Mdb_2=0$, and therefore the only configuration is F1s ending on the boundary. This is detailed in appendix \ref{app:FullSol} and we will see how to derive this condition from the 5d bulk topological action momentarily.

\vspace{0.1cm}
\noindent
\textbf{Mixed Anomaly and $\chi_{\text{SB}}$ from Holography.}
All things are now in place to see holographically the mixed anomaly \eqref{COanomaly} and $\chi_{\text{SB}}$ (\ref{BreakingStuff}). 
To do this, we need to study the rest of the topological couplings in the 5d bulk supergravity action. 
In particular we need to include the R-symmetry of the dual field theory, which is realized in terms of the $U(1)$-isometry (Reeb-vector) of the $T^{1,1}$ solution. This can be described by a $U(1)$ 1-form gauge field $A$, which enters the metric of $T^{1,1}$ as
$\d \beta \to \d \beta-A
$, where $\beta$ is the coordinate of the Hopf fiber of the $S^3$. 
The breaking by the ABJ anomaly to $\Z_{2M}$ is realized holographically by a St\"uckelberg coupling in the 5d consistent truncation. In addition, we argue that chiral symmetry breaking is consistent with the mixed 0-/1-form symmetry anomaly, which we derive from the 5d supergravity, and the KS-solution.
In the IR we also derive the TQFT proposed in \cite{Gaiotto:2014kfa} which matches the mixed anomaly. 

The additional 5d topological terms in the action are (see appendix \ref{sec:truncationembedding})
\be 
\label{eq:S5dtotal}
\ba
S_{\text{5d}} \supset 2 \pi  \int \frac{\cR}{2} |dc_0+2M A|^2- M^2 b_2^2 \, A +\frac{M}{2}b_2^2\, dc_0\,.
\ea\ee
The first term is the kinetic term for the axion. Since it contains two derivatives it is subleading when evaluated on the UV KS-solution, when $r$ is large, with respect to the topological terms. On the other hand its effect is important, since it realises the St\"uckelberg mechanism for the $U(1)_R$ gauge field $A$. The shift symmetry of the axion, $c_0 \sim c_0+2\pi$, is gauged by the $U(1)_R$ symmetry, so that the action is invariant under the non-linear transformation
\be 
A \to A+\d \alpha\,, \qquad c_0 \to c_0-2M \alpha\,.
\ee 
We can use this symmetry to completely gauge away the axion, leaving only a mass term for the gauge field. Fixing $c_0=0$, there is still a residual discrete symmetry generated by $\alpha \in \frac{\pi}{M} \Z$.
This is the direct way to identify the breaking of $U(1)_R \rightarrow \mathbb{Z}_{2M}^{(0)}$, as required by the ABJ anomaly.

The second term in \eqref{eq:S5dtotal} corresponds to the anomaly between the 0-form background $A$ for $\mathbb{Z}_{2M}^{(0)}$, $\oint A \in \frac{\mathbb{Z}}{2M}$, and $b_2$ for $\mathbb{Z}_{M}^{(1)}$, $\oint b_2 \in \frac{\mathbb{Z}}{M}$ 
\be 
\ba
\label{01Anomaly}
\mathcal{A}[b_2, A]& = - 2\pi M^2\int_{\cM_5}  b_2 \, b_2 \,  A \,,
\ea\ee 
which is a mixed 0-/1-form symmetry anomaly \cite{cupversuswedge}. As expected it does not depend on the energy scale, and therefore this term will survive in the IR. 
In the IR we expect the theory to be dual to a confining vacuum of $SU(M)$ SYM, so the $\mathbb{Z}_{M}^{(1)}$ should be unbroken, and this gapped phase should be described by a 4d TQFT.
Assuming $\mathcal{M}_5$ is spin, then $b_2^2$ is even. Since $A$ is a $\mathbb{Z}_{2M}$ background, \eqref{01Anomaly} does not become integral in general. It was proven in \cite{Cordova:2019bsd} that unless this term is integral there cannot be a 4d TQFT
 with $\Gamma^{(0)}=\mathbb{Z}_{2M}$ and $\Gamma^{(1)}=\mathbb{Z}_{M}$ symmetries in the IR that saturates all the anomalies of the theory in the UV. 
On the other hand, integrality of (\ref{01Anomaly}) and an unbroken $\mathbb{
Z}_M^{(1)}$ implies
\be
\oint A \in \frac{\mathbb{Z}}{2}:\qquad \mathbb{Z}_{2M} \ \rightarrow \ \mathbb{Z}_{2} \,,
\ee
implying chiral symmetry breaking in the IR vacuum of $SU(M)$ SYM. 
We stress that our analysis shows that the  presence of this topological coupling in the UV KS supergravity solution is already preempting and consistent with the chiral symmetry breaking in the gapped confining vacuum with $\mathbb{Z}_2^{(0)}$ and $\mathbb{Z}_{M}^{(1)}$ symmetries.

\vspace{0.1cm}
\noindent
\textbf{4d IR TQFT from Holography.}
Finally, we now turn to the IR description of the theory. 5d supergravity contains topological terms leading to the IR 4d  TQFT, which matches \eqref{01Anomaly}, and realises a spontaneous chiral symmetry breaking, $\mathbb{Z}_{2M} \ \rightarrow \ \mathbb{Z}_{2}$. 
From field theory,  the IR theory that  matches the anomalies of the UV gauge theory was proposed to be \cite{Gaiotto:2014kfa} 
\be\label{TQFTIR}
S_{\text{TQFT}_{\rm 4d}} = \int M \phi \left(d c_3+ \frac{M}{2}b_2^2\right) = \int M \phi F_4\,.
\ee
The $M$ vacua, labelled by $\langle e^{i \phi}\rangle =e^{\frac{2\pi i \ell}{M}}$, $\ell=0,1,\dots M-1$, are separated by domain walls (DWs) \cite{Gaiotto:2017tne}, $e^{i \oint c_3}$.  

The smooth IR gravity dual background is the deformed conifold solution, where $\tau \to 0$ (see appendix \ref{app:FullSol}). In this regime there is no hierarchy between the 5d bulk kinetic and topological terms, and the former need to be taken into account. Before the $S^2$ degenerates, the D5s source $C_6= \omega_3 \wedge c_3$, and in addition since  $F_7=\ast F_3$ we consider $dc_3=\frac{\cR}{2} \ast_5 (dc_0+2MA)$. Therefore, in the IR the dynamics of $c_0$ becomes relevant. Since $A$ corresponds to the true isometry of the IR KS-solution, that is a $\Z_2$ gauge field, $c_0$ does not shift under a gauge transformation of $A$. The 5d IR topological action becomes 
\be \label{eq:5dtopact}
S_{\text{5d}}' \supset  2 \pi \int  2M A dc_3 + dc_0 dc_3+\frac{M}{2}b_2^2\, dc_0 \,.
\ee
We first notice that the mixed anomaly between $\Gamma^{(0)}=\mathbb{Z}_{2M}$ and $\Gamma^{(1)}=\mathbb{Z}_M$ has disappeared. This is due to an additional topological term {$|c^{\Omega}(r)|^2 b_2^2 A $} \cite{Cassani:2010na, Berg:2005pd, Papadopoulos:2000gj}, which for the UV KS-solution depends on the UV scale $r_0$, but is constant  in the IR {$c^{\Omega} = M$}. This is consistent with anomaly matching since the IR theory has $\Gamma^{(0)}=\mathbb{Z}_{2}$, which is not anomalous on spin manifolds.
The third term is a total derivative, and varying by $c_0$ implies $M db_2=0$, in agreement with the equations of motion on the deformed conifold. 
When this condition is satisfied, the last two terms give rise to topological counterterms for the 4d theory living at the boundary. This implies that they are not anomalies, but rather the imprint of the TQFT  (\ref{TQFTIR}) in the IR, which is precisely obtained by identifying $c_0 \leftrightarrow  M\phi$ and evaluating these terms at the boundary. In particular, $c_0$ is related to the presence of DWs given by D5s wrapping $S^3$. These source $\int_B F_3$, where $B$ is the Poincar\'{e} dual cycle in the deformed conifold with $S^2$ boundary at infinity. This entails that {$\int_B F_3 \sim \int_{S^2} c_0 \omega_2 $}, and because of the presence of the D5 DWs, $c_0=\ell$ is quantized and corresponds to the number of D5s. The UV anomaly \eqref{01Anomaly} in the IR is realised by the action of $\Gamma^{(0)}=\mathbb{Z}_{2M}$,  $\ell \rightarrow \ell+1$, which is however not a symmetry of the IR vacuum. 

The IR theory proposed in \cite{Gaiotto:2014kfa} is furthermore invariant under 1-form symmetry transformation $B_2\rightarrow B_2 + d \lambda$, and this implies that $c_3\rightarrow c_3 - N B_2  \lambda  - \frac{N}{2} \lambda d \lambda$. The transformation of the 1-form symmetry enters in the shift of $c_3$, which signals the presence of a 3-group \cite{Seiberg:2018ntt, Gaiotto:2014kfa}. This is again supported by the string theory realisation of these DWs in terms of D5s wrapped on $S^3$ in the deformed conifold. The CS-action of the D5 is $ \mathcal{L}_{CS}=  \sum_p C_p \wedge e^{-B}$. The DWs extend in the 4d spacetime such that $F_4= \int_{S^3} d \mathcal{L}_{CS} = dc_3 + \frac{M}{2}B_2^2$, the 3-group follows from the gauge invariance of the world-volume action of the D5 and it is analogously consistent with gauge invariance of \eqref{eq:5dtopact}. 

\vspace{0.1cm}
\noindent
\textbf{Anomalies from Little String Theory.}
An alternative large $N$ limit of 4d $\mathcal{N}=1$ SYM is the solution in \cite{Maldacena:2000yy}, which is topologically twisted $S^2$-reduction of 6d little string theory (LST) on $N$ NS5-branes in IIB, which in the IR is 6d $(1,1)$ SYM.

We now derive the chiral symmetry breaking in 4d $\mathcal{N}=1$ SYM for gauge group $G$ from a twisted $S^2$-reduction of LST.
The anomaly polynomial of LSTs has a term \cite{Cordova:2020tij}
\begin{equation} \label{eq:6danpol}
	I_{8, \text{mixed}} \supset 2h^{\vee}_{G} c_2(R_{6d}) c_2(F_G) \,,
\end{equation}
where $R_{6d}$, $F_G$ are the $SU(2)_R$ and  gauge bundles, respectively.
We consider the decomposition 
$R_{6d}=(R \otimes K^{1/2}_{S^2}) \oplus (R \otimes K^{1/2}_{S^2})^{\vee}$, 
where $R$ is the 4d $U(1)_R$-bundle and $K_{S^2}$ is the canonical bundle on $S^2$. The twist is realized by
$c_2(R) = - \left(dA_R+ \frac{1}{2}c_1(K_{S^2})\right)^2$, 
where $A_R$ is associated to the 4d $U(1)_R$. Integrating \eqref{eq:6danpol} on $S^2$
\begin{equation}
	I_6 = \int_{S^2} I_{8, \text{mixed}} \supset 2 h^{\vee}_{G} dA_R \wedge c_2(F_G) \,.
\end{equation}
This is the ABJ anomaly of 4d $\mathcal{N}=1$ SYM, and it trivializes on a closed 6-manifold, $Y_6$, when $2 h^{\vee}_{G} dA_R=0$, i.e. when $A_R$ is in $\mathbb{Z}_{2 h^{\vee}_G}$. In fact, this corresponds to the breaking of the $U(1)_R \rightarrow \mathbb{Z}_{2 h^{\vee}_G}$ by the ABJ anomaly. 

We now also derive the mixed 0-/1-form  't Hooft anomaly between $\mathbb{Z}_{2 h^{\vee}_G}$ and 1-form symmetry $\Gamma^{(1)}=Z(G)$,  where $Z(G)$ is the center of the simply connected $G$.
Let $\mathcal{M}_5=\partial Y_6$ and consider the anomaly of the 4d theory, $I_5 \supset 2 h^{\vee}_{G} A_R \wedge c_2(F_G)$ such that $I_6=dI_5$. 
Activating a 1-form symmetry background $b_2\in H^2(\mathcal{M}_4,Z(G))$, makes the instanton density fractional, since it is evaluated in $G/Z(G)$ bundles, and
$I_5$ becomes 
\begin{equation} \label{eq:anfromLST}
	I_5 \supset 2 h^{\vee}_{G}  \alpha_G A_R \cup \mathfrak{P}(b_2) \,,
\end{equation}
where $\mathfrak{P}(b_2)$ is the Pontryagin square \cite{Cordova:2019uob} and for $SU(N)$ $\alpha_G = \frac{N-1}{2N}$. In particular, $A_R$ is not a $U(1)_R$ bundle, but rather a $\mathbb{Z}_{2 h^{\vee}_G}$ surviving subbundle, i.e. non-trivial configuration such that $2 h^{\vee}_{G} dA_R=0$ on a closed $Y_6$. Note that due to this, the anomaly does not trivialize anymore when $\alpha_G$ is fractional ($\oint \mathfrak{P}(b_2) \in \mathbb{Z}$). $A_R$ has then periodicity
$
\oint A_R \in \frac{\mathbb{Z}}{2 h^{\vee}_{G}}$
and from \eqref{eq:anfromLST} we have recovered the full mixed anomaly between the chiral symmetry {and the 1-form symmetry}, which we derived holographically in the previous section.


\begin{thebibliography}{42}
\bibitem [{oth()}]{otherbc}%
  \BibitemOpen
  \href@noop {} {\ }\bibinfo {note} {{{Note that there are other b.c. e.g.
  $\mathcal{C}$ Dirichlet/$b_2$ Neumann, which result in the same 1-form
  symmetry. For $\text{gcd}(M, N) = p q$ it is also possible to consider mixed
  b.c. which yield $\Gamma^{(1)}= \mathbb{Z}_p \times
  \mathbb{Z}_q$.}}}\BibitemShut {Stop}%
\end{thebibliography}

%


\onecolumngrid

\vspace{0.2cm}
\noindent
\textbf{Acknowledgements.}
We thank D. Cassani, O. Bergman, L. Bhardwaj, P. Benetti-Genolini, F. Bonetti, C. Closset, M. H\"{u}bner, C. Nu\~nez, L. Tizzano for discussions and/or comments on the draft.  
This work is supported by ERC grants 682608. SSN also acknowledges support through the Simons Foundation Collaboration on "Special Holonomy in Geometry, Analysis, and Physics", Award ID: 724073, Schafer-Nameki.


\appendix

\section{Full KS-solution, Deformed Conifold and 5d Topological Coupling in the IR}
\label{app:FullSol}

At small radial distances $r\rightarrow r_s$ of the KS-solution, higher curvature corrections cause the UV solution \eqref{eq:conmet} to break down. This regime, which we call the IR KS-solution, is sensitive to the deformation of the conifold induced by the $M$ D5-branes wrapping $S^2 \subset T^{1,1}$. The non-zero $F_3$ flux threading the $S^3$ prevents this cycle from shrinking to zero volume, whereas the $S^2$ collapses. Here the effective number of D3s is zero and the gauge theory dual is the IR regime of pure $\cN=1$ $SU(M)$ SYM. The warped, deformed conifold is parametrised by a new coordinate $\tau$, which, at large $\tau$, is related to $r$ by $r^2= 32^{-5/3}\epsilon^{4/3} e^{2\tau/3}$.
Near $\tau \rightarrow 0$ the metric is approximately $\mathbb{R}^{3,1}$ times the deformed conifold \cite{Klebanov:2000hb}. For the sake of illustration we include the shrinking $S^2$ in the degenerate metric 
\be\ba
ds_{10}^{2} &= c_1 \epsilon^{-4/3} (g_sMl_s^2)^{-1} \d \vec{x}^2 + c_2 g_s M l_s^2 ds_{6}^{2} \,, \\
ds_{6}^{2} &=\frac{1}{2}d\tau^{2} + \half (g^{5})^{2} + (g^{3})^{2} + (g^{4})^{2} + \frac{1}{4}\tau^{2}[(g^{1})^{2} + (g^{2})^{2}] \,,
\ea\ee
where $\{ g^i \}$ are the standard basis of 1-forms on $T^{1,1}$ \cite{Minasian:1999tt} and $c_i$ are numerical constants \cite{Klebanov:1998hh}. 
For $g_{s}M \ll 1$ the curvatures are small everywhere, even in the far IR, such that the supergravity approximation is always reliable. At $\tau=0$ the flux background has significant simplifications
\be 
F_5 = 0\,, \qquad H_3=0\,, \qquad F_3=\frac{l_s^2M}{2}g^5 \wedge g^3 \wedge g^4\,.
\ee 
Since the $S^2$ shrinks to zero radius as $\tau \rightarrow 0$, we modify our ansatz to ignore any fluctuations over the 2-sphere. This has the immediate consequence of removing all $F_5$ fluctuations, $\delta F_5=0$, as we would otherwise be unable to enforce self-duality. The ansatz is therefore $\delta H_3=h_3\,, \ \delta F_3=g_3$.
The Bianchi identities imply that $g_3=dc_2$, $h_3=db_2$ are closed and the equations of motion for $H_3$ and $F_3$ furnish two kinetic terms 
\be 
d \ast_5 h_3 = 0 \,, \qquad d \ast_5 g_3 =0 \,.
\ee
Contrary to the UV KS-solution, the kinetic terms are not suppressed in this regime. We get a single non-trivial contribution from the $F_5$ Bianchi identity implying that 
\be
Mdb_2 = 0\,.
\ee
We have no freedom in selecting boundary conditions here, the holographic duality picks out a confining vacuum and in particular a global form of the gauge group, namely the $SU(M)$ simply-connected group, with an unbroken electric 1-form symmetry $\Z_M^{(1)}$. From the bulk perspective, F1s can end on the boundary, indicating the presence of Wilson lines in the gauge theory. Furthermore, $M$ fundamental strings are screened by wrapped D5s that correspond to the "baryon vertex" in the 4d theory.

\section{Equations of Motion and 5d Supergravity} 

\label{app:eom}

In this appendix we derive the equations of motion of the 5d effective theory, obtained from compactifying 10d IIB supergravity on $T^{1,1}$. We isolate dominant topological couplings and determine an effective 5d action, which governs them. To identify the topological couplings, we expand field strengths $F_q$ along $\omega_p \in H^p(T^{1,1},\Z)$ as $F_q = \sum_p f_{q-p} \wedge \omega_p$, and insert these into the Type IIB equations 
\be 
\ba 
d H_3 &=0\,, \qquad &&d \ast_{10} H_3=-g_s^2 F_5 \wedge F_3\,,\\
d F_3 &=0\,, \qquad &&d \ast_{10} F_3= F_5 \wedge H_3\,,\\
d F_5 &= H_3 \wedge F_3\,, \qquad &&{\color{white} \d} \ast_{10} F_5=F_5\,.
\ea 
\ee 
The couplings obtained in this way can equally be thought of as embedded into some consistent truncation (e.g. \cite{Cassani:2010na}). We will find the following topological term in the 5d reduction in the KS-solution
\be 
\label{5dTop}
S_{\text{top}}=2 \pi \int_{\cM_5} b_2 \wedge \left( N dc_2 - M da_2 \right) \,.
\ee 
In this section we focus on the UV regime: large $r \sim r_0$, dual to the top of the cascade where both cycles $S^2 \times S^3 \subset T^{1,1}$ are non-degenerate. We expand the fluctuations along the volume forms $\omega_{2,3} \in H^\ast(T^{1,1},\Z)$ (see e.g. \cite{Herzog:2001xk} for conventions and an explicit parametrization)
\be 
\ba 
\delta F_3 &=g_3+\pi l_s^2 \omega_2 \wedge g_1\,, \qquad
\delta H_3 = h_3\,, \\
\delta F_5 &=  \pi l_s^2\omega_2 \wedge f_3 + \frac{2\pi l_s^2}{6}\cR \omega_3 \wedge \ast f_3 \,.
\ea 
\ee 
Here, $h_3,  g_{1,3}, f_3$ are all external fields and in this appendix we restore pre-factors for completeness. Self-duality of $\delta F_5$ implies a choice of frame: we can fix one expansion component in terms of the other. We use the 3-form piece, since the operators of the boundary 1-form symmetry are manifest in this frame.
The Bianchi identities for $H_3, F_3$ imply that the corresponding 5d fields are closed, so we write $h_3 = db_2,  g_3 = dc_2,  g_1 = dc_0$. We interpret $c_0$ as an axion, whereas $b_2, c_2$ couple to F1s and D1s, respectively.
The Bianchi identity for $\delta F_5$ implies that $f_3$ is not closed 
\be
d  f_3=  \d \cL \wedge g_3 + h_3 \wedge g_1 \,.
\ee
As such, we shift the field to obtain closure and define a new gauge potential $da_2=f_3 - \cL dc_2 - b_2dc_0$, {which couples to D3s wrapping $S^2 \subset T^{1,1}$}. The 5d equations of motion are
\be 
\ba 
\label{eq:bianchi5}
d (  \cR \ast_5  f_3 ) &= \frac{3}{2\pi} M {db_2} \\
d ( \cR^5 \ast_5 {db_2} )&=-27\pi l_s^4 g_s^2 ( \cK {dc_2}-M  f_3+\frac{2\pi}{3} \cR \ast_5  f_3 \wedge {dc_0}  ) \\
d ( \cR^5 \ast_5 {dc_2} )&=27\pi l_s^4 ( \cK {db_2}+\frac{2\pi}{3} \cR \ast_5  f_3 \wedge \d \cL )\\
d ( \cR \ast_5 dc_0 )&=-  \cR \ast_5 f_3 \wedge db_2\,.
\ea 
\ee 
From these equations of motion we extract leading topological contributions
\be
Ndb_2=0\,, \qquad  Mdb_2 = 0\,, \qquad  Ndc_2 - Mda_2= 0 \,,
\ee
where we ignore $c_0$, which can be gauged away via a St\"uckelberg mechanism.
From these we construct the action \eqref{eq:truetop}.

\section{5d Consistent Truncation}
\label{sec:truncationembedding}

An important component of the analysis presented in this paper is the consistent truncation of IIB supergravity to 5d for conifold solutions. In \cite{Cassani:2010na} such a consistent truncation was found which encompasses both the UV and IR KS-solutions, and where we show, the holographic realization of the ABJ anomaly and the mixed 0-/1-form symmetry anomaly are both manifest.
In what proceeds, we present the map  required to translate between our work and their notation.
The KS flux background is parametrised as
\be 
\label{eq:Cassflux}
F_3=q \Phi \wedge \eta\,, \qquad B_2=b^\Phi \Phi\,, \qquad F_5=-(k-qb^\Phi) \Phi \wedge \Phi \wedge \eta\,,
\ee 
where $\Phi, \eta$ are left-invariant forms on $T^{1,1}$. They are related to the volume forms of $S^2$ and $S^3$ as $\Phi = \frac{1}{3}\omega_2\,, \ \Phi \wedge \eta = -\frac{1}{9} \omega_3$. 
We rescale the IIB fields by
\be
\label{eq:rescaleF} 
F_3 \to -\frac{9l_s^2}{2} F_3\,,  \qquad B_2 \to -3\pi l_s^2 B_2\,, \qquad F_5 \to \frac{27 \pi l_2^4}{2} F_5\,, 
\ee 
which ensures that the background is quantised as
\be\ba
\int_{S^3} \frac{F_3}{(2 \pi l_s)^2}=q \in \Z \,, \qquad
\int_{T^{1,1}} \frac{F_5}{(2 \pi l_s)^4}= k \in \Z\,,
\ea\ee
and furthermore gives the identifications 
\be 
q = M\,, \qquad b^\Phi = -\cL\,, \qquad k = N\,.
\ee
Notice that the rescalings are consistent: they give rise to the same factor on either side of the Bianchi identity for $F_5$.
These normalisations also imply that we should identify the fluctuation $d c^\Phi = -\frac{2\pi}{3} d c_0$. Finally, we rescale the $U(1)$ gauge field in \cite{Cassani:2010na}  $A \to \frac{4\pi}{3} A$ so that it is normalised as in \cite{Herzog:2002ih}. 
The 5d topological couplings obtained in \cite{Cassani:2010na} are
\be 
\cL_{5d} = \cR |g_1^{\Phi}|^2-\half f_2^{\Phi} \wedge \lb -q b_2 \wedge A+b_2 \wedge D c^\Phi \rb\,,
\qquad
f_2^\Phi \supset qb_2\,, \quad g_1^\Phi \supset Dc^\Phi\,, \quad Dc^\Phi = dc^\Phi -qA\,.
\ee
Using the map detailed above gives the action \eqref{eq:S5dtotal}.



\end{document}